\documentclass[prl,twocolumn,showpacs,showkeys,preprintnumbers,amsmath,amssymb]{revtex4}

\usepackage{graphicx}
\usepackage{dcolumn}
\usepackage{bm}

\def\duzomniejsze{<\kern-.7mm<}
\def\duzowieksze{>\kern-.7mm>}

\def\textbf#1{{\bf #1}}
\def\beq{\begin{equation}}
\def\eeq{\end{equation}}
\def\be{\begin{equation}}
\def\ee{\end{equation}}
\def\ben{\begin{eqnarray}}
\def\een{\end{eqnarray}}
\def\beqa{\begin{eqnarray}}
\def\eeqa{\end{eqnarray}}
\def\eea{\end{array}}
\def\bea{\begin{array}}
\newcommand{\bei}{\begin{itemize}}
\newcommand{\eei}{\end{itemize}}
\newcommand{\bee}{\begin{enumerate}}
\newcommand{\eee}{\end{enumerate}}

\usepackage{xcolor}
\newcounter{mnotecount}[section]
\renewcommand{\themnotecount}{\thesection.\arabic{mnotecount}}
\newcommand{\mnote}[1]
{\protect{\stepcounter{mnotecount}}$^{\mbox{\footnotesize
$
\bullet$\themnotecount}}$ \marginpar{
\raggedright\tiny\em
$\!\!\!\!\!\!\,\bullet$\themnotecount: #1} }

\usepackage{soul}

\begin{document}



\title{The curious case of operators with spectral density increasing as $\Omega(E)\sim e^{\,\mathrm{Const.}\, E^2}$}%

\author{Erik Aurell}
\email{eaurell@kth.se}
\affiliation{KTH -- Royal Institute of Technology, AlbaNova University Center, SE-106 91 Stockholm, Sweden}%

\author{Satya N. Majumdar}
\email{satyanarayan.majumdar@cnrs.fr}
\affiliation{LPTMS, CNRS, Univ. Paris-Sud, Universit\'e Paris-Saclay, 91405 Orsay, France}

\date{\today}

\begin{abstract}
Motivated by a putative model of black holes as quantum objects we consider what types of operators would have a corresponding spectrum. We find that there are indeed such operators, but of a rather unusual types, and for which the wave functions are only barely localized. We point out a tension between such almost delocalized states and black holes as compact objects. 
\end{abstract}

\maketitle

{\it Introduction:} 
According to Bekenstein and Hawking, a black hole 
has an (non-dimensional) entropy $\frac{1}{4}\frac{A}{\ell_P^2}$ where $A$ is the area of the black hole horizon, and $\ell_P\approx 
1.6 \times 10^{-35}\, \mathrm{m}$ is the Planck length. For an electrically neutral spherically symmetric (Schwarzschild) black hole this is also $4\pi\,\frac{M^2}{m_P^2}$ where 
$M$ is the black hole mass, and 
$m_P\approx 
2.2\times 10^{-8}\, \mathrm{kg}$ is the Planck mass.
If Boltzmann's formula holds for this entropy,
for a solar mass black hole ($M\approx 2.0\times 10^{\,30}\,\mathrm{kg}$), it would correspond to about $e^{10^{77}}$ activated degrees of freedom. It is a major unsolved problem of modern fundamental physics what these degrees of freedom are. 

Let us note that the estimate of the black hole entropy can be arrived at in several ways. The most immediate is to use the classical thermodynamic formula $\frac{1}{T}=\frac{\partial S}{\partial E}$ in a setting where $E=Mc^2$ is the energy and $T$ is the Hawking temperature, determined by Bogolyubov transformation between ingoing and outgoing states of elementary particles propagating in the curved space-time around the black hole \cite{Hawking1974,Hawking1975,Wald75,wald1994quantum}. It makes no reference to anything happening inside the black hole. The second is to estimate the number of quantum states of matter that could collapse and give rise to a black hole with given macroscopic black hole parameters (mass, electric charge, angular momentum). Such an estimate was first carried out by Bekenstein in the very first paper on black hole entropy
\cite{Bekenstein1973}, and later refined by Mukhanov in \cite{Mukhanov2003}. From the viewpoint of statistical mechanics this is a coarse-grained entropy (coarse-graining by the outside observer only knowing about mass, electric charge and angular momentum). 
As it is based on ignorance of what goes on the inside it also makes no statement on what the excited degrees of freedom of the black hole may be. 

The number of hypotheses that have been put forward to describe black holes, and which may give a statistical mechanics (or other) interpretation of black hole entropy, is too vast to be fully cited here; we refer to well-known reviews \cite{Bekenstein:2008,
Page2005,HossenfelderSmolin2010,Harlow,Marolf} and recent high-profile contributions
\cite{Almheiri2019,Almheiri2020,Penington2020,Almeiri21,Penington2022,Balasubramanian2024b,Balasubramanian2024a}, and papers cited therein.
However, if such an interpretation exists, and if it is based on quantum mechanics, then a black hole with mass about $M$ would, as a quantum object, have a number of states of approximately that mass, and the black hole entropy would be the logarithm of the density of states (ignoring the other two macroscopic parameters, electric charge and angular momentum). 

Mukhanov and Bekenstein 
\cite{Mukhanov1986, Bekenstein1995, Mukhanov2018}, see also \cite{BekensteinMeisels77}, hence proposed that the masses of black holes are quantised in multiples of the Planck mass 
(as $M_n\sim \sqrt{n}\, m_P$), each state being exponentially degenerate
($D_n\sim e^{\alpha n}$). Adjusting suitably the constants one can thus arrive at an interpretation of black hole entropy as $S(M)=\log D_{n}$ where $n=\frac{M^2}{m_P^2}$.  
What has not been discussed previously is if and when such spectra of operators can actually occur in reasonable mathematical models. To do so is our goal here.

\textit{A simple model:}
We consider a quantum gas of $N$ non-interacting bosons.
Since the bosons are non-interacting, this many-body system can be described
in terms of the energy levels of a single particle and the associated occupation numbers.
More precisely, 
let $\{\epsilon_i\}$ denote the single particle spectrum. For convenience, we choose 
$\epsilon_i\ge 0$ without
any loss of generality. Then a microscopic state
of the system can be labelled by $\{n_i\}$, where $n_i\ge 0$ is the occupation number
of the $i$-th single particle level. The microcanonical partition function
$\Omega(E)$, associated to a total energy $E$ of the many-body system  can then 
be simply expressed as 
\begin{equation}
\Omega(E)= \sum_{n_i\ge 0} \delta\left(E- \sum_{i} n_i\right)\, .
\label{micro.2}
\end{equation}
This model of noninteracting bosons has appeared in many contexts, most recently in the context of cold atom experiments. It has numerous applications in many domains of physics and mathematics, including 
even in number theory. For example, how many ways an integer $E$ can be partitioned, i.e., expressed as a sum of smaller integers?  Hardy and Ramanujam showed that $\Omega(E)\sim \exp\left[\pi\, \sqrt{2/3}\, \sqrt{E}\right]$ for large E~\cite{HR1918}. This problem and many of its
generalizations concerning combinatorial aspects of the integer partition problem have been studied in the literature~\cite{EL1941,AK1946,AA1951,HT1954}. It turns out that this 
integer partition problem and its variations
can be mapped into the model of noninteracting bosons which helped explore many interesting questions both in physics and mathematics~\cite{TMB2004,CLM2007,CMO2007,CMOS2007,EMULR2024,PBM2024}. For example, using this mapping a very interesting
connection was found between the integer partition problem 
and the extreme value statistics~\cite{CLM2007}.

To compute $\Omega(E)$ in \eqref{micro.2}, we first take its Laplace transform with
respect to $E$ and then sum over all $n_i\ge 0$. This gives
\begin{equation}
Z(\beta)= \int _0^{\infty} e^{-\beta E}\, \Omega(E)\, dE= 
\prod_{i} \frac{1}{1- e^{-\beta\, 
\epsilon_i}}\, .
\label{can_pf.1}
\end{equation}
Inverting formally this Laplace transform one gets
\begin{equation}
\Omega(E)= \int_{\Gamma} d\beta\, 
e^{\beta E- \sum_{i} \ln \left(1-e^{-\beta\, \epsilon_i}\right)}\, = \int_{\Gamma} 
d\beta\, e^{F(\beta)},
\label{micro.3}
\end{equation}
where $\Gamma$ denotes the Bromwich contour in the complex-$\beta$ plane
and we denoted
\begin{equation}
F(\beta)= \beta\, E-  \sum_{i} \ln \left(1-e^{-\beta\, \epsilon_i}\right)\, .
\label{FB.1}
\end{equation}
To evaluate this integral in \eqref{micro.3} we employ the
standard saddle point method by assuming $F(\beta)$ to be large 
(to be justified aposteriori). Minimizing $F(\beta)$ with respect to $\beta$,
i.e., setting  $F'(\beta)=0$ at $\beta=\beta^*$ leads to the saddle point equation
\begin{equation}
E= \sum_{i} \frac{\epsilon_i}{e^{\beta^*\, \epsilon_i}-1}\, .
\label{sp.1}
\end{equation}
Note that $\langle n_i\rangle=
1/\left(e^{\beta^* \epsilon_i}-1\right)$ in Eq. (\ref{sp.1}) is simply the Bose factor
associated to the canonical partition function of the gas at an inverse temperature $\beta^*$.
For a given $E$, one needs to first solve the saddle point equation (\ref{sp.1})
to express $\beta^*$ in terms of $E$
and then evaluate $F(\beta^*)$. Consequently, $\Omega(E)$ can be estimated (up to pre-exponential factors) as
\begin{equation}
\Omega(E) \approx e^{F\left(\beta^*(E)\right)}\, .
\label{sp.2}
\end{equation}

For large $E$, one can approximate the discrete sum in Eq. (\ref{sp.1}) by an integral
over the single particle energy $\epsilon$
\begin{equation}
E\approx \int_0^E \frac{\epsilon\, \rho(\epsilon)\, d\epsilon}{e^{\beta^*\, \epsilon}-1}\, ,
\label{sp.3}
\end{equation}
where $\rho(\epsilon)$ denotes the density of energy states in the single particle problem.
We have deliberately kept the upper limit $E$ since one can not occupy levels 
higher than $E$ if the total energy is $E$. Clearly, if $\rho(\epsilon)$ grows slower than
$e^{\epsilon}$ for large $\epsilon$, to leading order for large $E$, one can replace
the upper cutoff in the integral in \eqref{sp.3} by $\infty$. For example, 
if $\rho(\epsilon)\sim \epsilon^\alpha$ as $\epsilon\to \infty$ with $\alpha>0$, one estimates
\begin{equation}
E\approx \int_0^{\infty} \frac{\epsilon^{\alpha+1}\, 
d\epsilon}{e^{\beta^*\, \epsilon}-1}
\sim \left[\beta^*\right]^{-(\alpha+2)}\, ,
\label{sp_pl.1}
\end{equation}
implying $\beta^*(E)\sim E^{-1/(\alpha+2)}$ for large $E$. This then leads to
$F(\beta^*(E))\sim E^{(\alpha+1)/(\alpha+2)}$ and hence, for large $E$,
\begin{equation}
\Omega(E) \sim \exp\left[E^{(\alpha+1)/(\alpha+2)}\right]\, 
\quad {\rm for} \quad \rho(\epsilon)\sim \epsilon^{\alpha}\, .
\label{micro_pl.1}
\end{equation}
For instance, for the integer partition problem, one can show that $\alpha=0$~\cite{CLM2007}, leading to the Hardy-Ramanujam result $\Omega(E)\sim \exp[\sqrt{E}]$~\cite{HR1918}.
Since the growth exponent $(\alpha+1)(\alpha+2)<1$, an algebraically growing 
$\rho(\epsilon)$ can not give rise to an $\Omega(E)$ growing faster than
exponential for large $E$, and in particular it can not
give rise to an $\Omega\sim e^{C_0\, E^2}$ as required for a black hole.
This remains true for any $\rho(\epsilon)$ that grows slower than an exponential for large
$\epsilon$.

\textit{A high energy condensation scenario that leads to the right spectrum:}
To reproduce an $\Omega(E)$ that grows faster than exponential for large $E$, we
need a $\rho(\epsilon)$ that also grows faster than exponential for large $\epsilon$.
In that case, we need to keep the upper cutoff $E$ in the integral in \eqref{sp.3} (as otherwise
the integral will be divergent)
and the integral will be completely dominated by the contribution coming from
the vicinity of this upper cutoff. Indeed, to leading order in $E$, one can them approximate
the integral in \eqref{sp.3} by 
\begin{equation}
E \sim  \frac{E\, \rho(E)}{e^{\beta^* E}-1} \approx E \rho(E) e^{-\beta^*\, E}\, ,
\label{sp_app.1}
\end{equation}
leading to 
\begin{equation}
\beta^*(E) \sim \frac{\ln\left[\rho(E)\right]}{ E}\, .
\label{sp.5}
\end{equation}
Consequently, from \eqref{FB.1} and \eqref{sp.2} it follows that
\begin{equation}
\Omega(E) \sim \rho(E)\, .
\label{sp_final.1}
\end{equation}
This result has a very simple and nice physical picture associated to it.
This is an example of a `high energy condensation' where one or few particles
at a very high single particle level 
need to occupy a macroscopically large fraction of the total energy $E$. Thus
most of the energy is carried by this high energy condensate consisting of
one or few particles. This is a counterpart to the standard Bose-Einstein condensation
that occurs in Bose gases at low energy and low density. Thus, this high energy condensation
is the physical mechanism that can produce an $\Omega(E)$ growing faster than 
an exponential for large $E$. According to this mechanism and the result
in \eqref{sp_final.1}, if we want to
reproduce the black hole result $\Omega(E)\sim \exp[C_0 E^2]$ for large $E$ within
this simple non-interacting quantum Bose gas model, we need to
have a single particle quantum spectrum whose density of states $\rho(\epsilon)$
grows also as $\rho(\epsilon)\sim \exp[C_0\, \epsilon^2]$ for large $\epsilon$.

\textit{The quantum potential that leads to $\rho(\epsilon) \sim e^{C_0\, \epsilon^2} $:}
As a warm up, let us first discuss two well known 
simple examples: (i)
a $d$-dimensional harmonic oscillator with $V(r)= r^2$ and (ii) a $d$-dimensional
box of size $L^d$. In case (i), the single particle energies are given by
$\epsilon= \sum_{j}^d m_j$ (up to a constant and in suitable units) where
$m_j=0,1,2\ldots$ are the quantum numbers. Consequently, 
\begin{equation}
\rho(\epsilon) = \sum_{m_j=0}^{\infty}\, \delta\left (\epsilon- \sum_{j=1}^d m_j\right)\, .
\label{rho_ho.1}
\end{equation}
Taking a Laplace transform with respect to $\epsilon$ and carrying out the sums one finds
trivially
\begin{equation}
\tilde{\rho}(s)= \int_0^{\infty} \rho(\epsilon)\, e^{-s\,\epsilon}\, d\epsilon \sim 
s^{-d} \quad {\rm as}\quad s\to 0\, ,
\label{rho_ho.lt}
\end{equation} 
leading to the well known density of states for a $d$-dimensional harmonic oscillator
\begin{equation}
\rho(\epsilon) \sim \epsilon^{d-1} \quad {\rm as} \quad \epsilon\to \infty\, .
\label{rho_ho_final}
\end{equation}
The example (ii), i.e., the $d$-dimensional box can be solved in a similar way by noting
that in this case $\epsilon= \sum_{j=1}^d m_j^2$ (in suitable units where $L=1$).
Proceeding as in case (i), it is easy to show that for the $d$-dimensional box
\begin{equation}
\rho(\epsilon) \sim \epsilon^{d/2-1} 
\quad {\rm as} \quad \epsilon\to \infty\, . 
\label{rho_box_final}
\end{equation}
In both examples, $\rho(\epsilon)$ grows algebraically for large $\epsilon$.

To estimate $\rho(\epsilon)$ for a general confining central potential $V(r)$ in $d$ dimensions, we can use
the semi-classical Bohr-Sommerfeld quantization formula which gives an accurate
estimate for large $\epsilon$. To make this estimate, it is convenient to work
with the cumulative density of states up to level $\epsilon$, i.e., 
${\cal N} (\epsilon)= \int_0^{\epsilon} \rho(\epsilon')\, d\epsilon'$. Then the integrated density of states scales 
as ${\cal N}(\epsilon)\sim \left[n(\epsilon)\right]^d$ in $d$-dimensions, where $n(\epsilon)$
is the quantum number (radial) associated to the energy level $\epsilon$. The latter
can be estimated from the Bohr-Sommerfeld quantization rule (again in appropriate units)
\begin{equation}
\int_{0}^{\infty} \sqrt{\epsilon- V(r)}\, dr \approx n(\epsilon)\sim  
\left[{\cal N}(\epsilon)\right]^{1/d} \, .
\label{BS.1}
\end{equation}
It is easy to check that this general result reproduces correctly the exact estimate
in Eq. (\ref{rho_ho_final}) for the harmonic oscillator case $V(r)=r^2$, as well
as the result in Eq. (\ref{rho_box_final} ) for the box case where $V(r)=0$.
Now, we want $\rho(\epsilon)\sim e^{C_0\, \epsilon^2}$. This means, up to pre-exponential factors,
${\cal N} (\epsilon) \sim e^{C_0 \epsilon^2}$. Substituting this on the right hand side (rhs) of
Eq. (\ref{BS.1}) gives for large $\epsilon$
\begin{equation}
\int_{0}^{\infty} \sqrt{\epsilon- V(r)}\, dr \approx e^{C_0\, \epsilon^2/d}\, .
\label{dos_re.1}
\end{equation}
A little inspection shows that a potential $V(r)$ that reproduces the rhs of \eqref{dos_re.1}
is of the form
\begin{equation}
V(r) \sim B\, \sqrt{\ln r} \, \quad {\rm as}\quad r\to \infty\, ,
\label{VR.1}
\end{equation}
where the prefactor $B$ can be determined as follows. We substitute this ansatz on the
left hand side of \eqref{dos_re.1} and make a change of variable $B\, \sqrt{\ln r}= \epsilon\, y$.
Consequently the left hand side (lhs) of \eqref{dos_re.1} reads
\begin{equation}
\int_{0}^{e^{\epsilon^2/B^2}} \sqrt{\epsilon- V(r)}\, dr = \frac{2\, \epsilon^{5/2}}{B^2}\, \int_0^1 \sqrt{1-y}\ y\, e^{\epsilon^2 y^2/B^2}\, dy\, .
\label{VR.2}
\end{equation}
For large $\epsilon$, the dominant contribution to the integral over $y$ comes from the vicinity of $y=1$ and one can easily show that up to inconsequential factors
\begin{equation}
\int_{0}^{e^{\epsilon^2/B^2}} \sqrt{\epsilon- V(r)}\, dr \approx e^{\epsilon^2/B^2} \quad {\rm as} \quad 
\epsilon \to \infty\, .
\label{VR.3}
\end{equation}
Comparing it to the rhs of \eqref{dos_re.1} fixes the prefactor $B= \sqrt{d/C_0}$ uniquely.
Hence, the required potential to produce $\Omega(E)\sim e^{C_0\, E^2}$ for large $E$ is
given by
\begin{equation}
V(r) \approx \sqrt{\frac{d}{C_0}\, \ln r} \quad \rm {as} \quad r\to \infty\, .
\label{VR.4}
\end{equation}

\textit{Eigenfunctions in a potential $\alpha\sqrt{\ln |x|}$:}
The potential in Eq. (\ref{VR.4}) grows extremely slowly with distance
from the origin. 
It may be questioned if a potential increasing as slowly
as the square root of the logarithm is even confining and
hence has discrete eigenvalues. In other words, are the
eigenfunctions sufficiently localised and square integrable?
We now show that they are indeed localised and for simplicity,
we just show this in $d=1$. Generalization to higher dimensions is straightforward.
We then consider a single quantum particle in one dimension in a potential
$V(x)=\alpha\, \sqrt{\ln |x|}$ with $\alpha>0$.
The eigenfunction $\psi_{\epsilon}(x)$ associated to an eigenvalue $\epsilon$ 
satisfy the Schr\"odinger equation
\begin{equation}
-\frac{\hbar^2}{2m}\, \frac{d^2\psi_{\epsilon}(x)}{dx^2} + 
V(x)\, \psi_{\epsilon}(x)= \epsilon\, \psi_{\epsilon}(x)\, .
\label{SE.1}
\end{equation}
where $V(x)=\alpha\sqrt{\ln |x|}$.
According to the WKB approximation, a good estimate of the symmetric
wave function for $x>0$ can be expressed as
\begin{equation}
\psi_{\epsilon}(x) \approx C_+\, \frac{\exp\left[\frac{i}{\hbar}\, \sqrt{2m}\,
\int_0^{x} \sqrt{\epsilon-V(x')}\, dx'\right]}{\sqrt{2m\, (\epsilon-V(x))}}\, ,
\label{WKB.1}
\end{equation}
where $C_+$ is a constant.
The real
line is thus divided into a "classically
allowed" region where $V(x)<\epsilon$,
and a "classically forbidden" region where $V(x)>\epsilon$. 
The two regions are separated by the "turning point" $x_c$ given by
$\epsilon=\alpha\sqrt{\ln |x_c|}$, i.e., 
\begin{equation}
x_c=\exp(\epsilon^2/\alpha^2)\, .
\label{eq:x_c}
\end{equation}
We are interested in the large $x$ regime, i.e., when $x>x_c$ (classically forbidden region) 
such that $V(x)>\epsilon$.
If $V(x)\gg \epsilon$, we see
from Eq. (\ref{WKB.1}) that to leading order for large $x$, the wavefunction 
(up to inconsequential prefactors) decays as
\begin{equation}
\psi_{\epsilon}(x) \propto  {\exp\left[-\frac{\sqrt{2m}}{\hbar}\,
\int_0^{x} \sqrt{V(x')}\, dx'\right]}, \, .
\label{WKB.2}
\end{equation}
Substituting $V9x)=\alpha\, \sqrt{\ln |x|}$ in \eqref{WKB.2}, we get for $x\gg x_c$
\begin{equation}
\psi_{\epsilon}(x) \propto 
\exp\left[-\frac{\sqrt{2m\, \alpha}}{\hbar}\, (\ln x)^{1/4}\, x \right] \, ,
\label{WKB.4}
\end{equation}
which, of course, is square integrable.
Hence the potential $V(x)=\alpha\, \sqrt{\ln |x|}$
is indeed "confining", and has discrete eigenvalues.


\textit{Considerations of interactions:} So far we have considered black holes as putative quantum objects and discussed which
confining potentials could give rise to spectra that in turn would be similar to
Bekenstein-Hawking entropy. We have shown that there exist very shallow potentials
growing as $\sqrt{\ln r}$ with the distance that gives rise to such spectra via
a high energy condensation mechanism.
However, such shallow potentials lead to states with
wavefunctions that, although
square integrable, have a large spatial extension which contradicts the idea of 
compact black holes. 
One way out could be to relax the ansatz of noninteracting bosons, and to postulate that self-interactions could at the same time lead to the right spectral degeneracy and the right spatial extent, both considered as functions of black hole mass. Such a scenario was indeed suggested some time ago by Dvali and Gomez\cite{dvali2011blackholesquantumnportrait}, see also \cite{Dvali}. This approach has however so far not been systematically investigated from the point of view of analysis of operators.
Within our approach, since the relevant states that contribute to the superexponential density of states
via the high energy condensation mechanism occur at very high energies, interactions are unlikely to play
any role there. 

\textit{The size of internal space:}
The above discussion supposes that the internal space of a black hole
can be assimilated to an object like an atom, figuratively
that the Schwarzschild radius of a black hole is analogous
to the Bohr radius. This does not have to be the case, as the geometry of a black hole interior may be qualitatively different
from the domain of ordinary space occluded by the black hole.
Indeed, the mathematical structure of space-time inside a (classical) black hole is still an open problem and the focus of a substantial literature, see \textit{e.g.}
\cite{Poisson1989,Bonnanno1995,Christodoulou1999,Dafermos2005,dafermos2017interiordynamicalvacuumblack}.
Recently the possibility that the (classical) gravitational dynamics inside a black hole being chaotic has been discussed
\cite{DeClerck2024,Bueno2024,Caceres2024} in specific models, in which case analytical solutions are likely to remain elusive.
The (semiclassical) quantum structure of black hole interiors is also actively investigated, see \textit{e.g.}
\cite{Zilberman2022,Carballo-Rubio2024}.

The internal volume of black hole might hence be large, often referred to as a 'bag-of-gold' scenario\cite{Wheeler1963}, and, if so, may accommodate many internal states without running into the conceptual problem outlined above. It is however not known why the internal volume should then always be of just the amount of vastness required to match Bekenstein-Hawking entropy, and in the recent literature the issue is raised that it may even be too large; one then has to explain why black hole entropy is not even larger\cite{Chakravarty2021}.
If a (classical) singularity is taken seriously as an end point of (classical) evolution in the black hole interior, one is further presumably not limited to internal spaces of finite dimension, but could also contemplate infinite-dimensional internal spaces, for instance the locally tree-like structures favored in disordered systems theory\cite{MezardMontanari2009,FarBeyond},
where the density of local ground states is known to be an exponentially increasing function of energy \cite{MezardParisi2003} (eq.~20).
In the context of 
(\ref{micro_pl.1})
this corresponds to the same scaling as in the limit $\alpha\to \infty$.
However, it is not known how to get to a spectral function increasing as exponential of a square in such a scenario. 

\textit{Conclusion:} Motivated by the spectral degeneracy of a black hole as a quantum object we have
considered the general question which hermitian operators can have spectral density growing with energy E as
$\sim \exp[{\mathrm Const.}\,  E^2]$. We have shown that typically such quickly
increasing degeneracy and concomitant cumulant state
counting function lead to the phenomenon of high energy condensation. We have further shown an explicit
family of Hamiltonian operators in finite dimensions (in-
cluding one dimension) with these properties. As we also
show, these examples lead to very extended states, and
therefore cannot be reasonable models of quantum black
holes, as their support would extend (very) far outside
the Schwarzschild radius. We have briefly considered these conclusions
in the perspective of interacting bosons, and have pointed out
that to change the conclusions qualitatively the interactions would have to shrink the spatial extent of the
relevant states in a major way. We have also commented on the
possibility of different geometries inside black holes which
can correspond to these highly degenerate spectra. On
a general level, our work serves to highlight what an unusual and indeed extreme mathematical object 
a quantum system must be to have the Bekenstein-Hawking-Mukhanov
energy spectrum.

\section*{Acknowledgement}
EA thanks Prof V Mukhanov for discussions around the 'bag-of-gold' scenario. SNM acknowledges ANR Grant No. ANR- 23-CE30-0020-01 EDIPS.

\bibliography{GaussianHawking,refs-SI}

\begin{thebibliography}{50}
\expandafter\ifx\csname natexlab\endcsname\relax\def\natexlab#1{#1}\fi
\expandafter\ifx\csname bibnamefont\endcsname\relax
  \def\bibnamefont#1{#1}\fi
\expandafter\ifx\csname bibfnamefont\endcsname\relax
  \def\bibfnamefont#1{#1}\fi
\expandafter\ifx\csname citenamefont\endcsname\relax
  \def\citenamefont#1{#1}\fi
\expandafter\ifx\csname url\endcsname\relax
  \def\url#1{\texttt{#1}}\fi
\expandafter\ifx\csname urlprefix\endcsname\relax\def\urlprefix{URL }\fi
\providecommand{\bibinfo}[2]{#2}
\providecommand{\eprint}[2][]{\url{#2}}

\bibitem[{\citenamefont{Hawking}(1974)}]{Hawking1974}
\bibinfo{author}{\bibfnamefont{S.~W.} \bibnamefont{Hawking}}, \bibinfo{journal}{Nature} \textbf{\bibinfo{volume}{248}}, \bibinfo{pages}{30} (\bibinfo{year}{1974}).

\bibitem[{\citenamefont{Hawking}(1975)}]{Hawking1975}
\bibinfo{author}{\bibfnamefont{S.~W.} \bibnamefont{Hawking}}, \bibinfo{journal}{Communications in Mathematical Physics} \textbf{\bibinfo{volume}{45}}, \bibinfo{pages}{199} (\bibinfo{year}{1975}).

\bibitem[{\citenamefont{Wald}(1975)}]{Wald75}
\bibinfo{author}{\bibfnamefont{R.~M.} \bibnamefont{Wald}}, \bibinfo{journal}{Communications in Mathematical Physics} \textbf{\bibinfo{volume}{45}}, \bibinfo{pages}{9} (\bibinfo{year}{1975}).

\bibitem[{\citenamefont{Wald}(1994)}]{wald1994quantum}
\bibinfo{author}{\bibfnamefont{R.~M.} \bibnamefont{Wald}}, \emph{\bibinfo{title}{Quantum field theory in curved spacetime and black hole thermodynamics}} (\bibinfo{publisher}{University of Chicago press}, \bibinfo{year}{1994}).

\bibitem[{\citenamefont{Bekenstein}(1973)}]{Bekenstein1973}
\bibinfo{author}{\bibfnamefont{J.~D.} \bibnamefont{Bekenstein}}, \bibinfo{journal}{Phys. Rev. D} \textbf{\bibinfo{volume}{7}}, \bibinfo{pages}{2333} (\bibinfo{year}{1973}), \urlprefix\url{https://link.aps.org/doi/10.1103/PhysRevD.7.2333}.

\bibitem[{\citenamefont{Mukhanov}(2003)}]{Mukhanov2003}
\bibinfo{author}{\bibfnamefont{V.~F.} \bibnamefont{Mukhanov}}, \bibinfo{journal}{Foundations of Physics} \textbf{\bibinfo{volume}{33}}, \bibinfo{pages}{271} (\bibinfo{year}{2003}).

\bibitem[{\citenamefont{Bekenstein}(2008)}]{Bekenstein:2008}
\bibinfo{author}{\bibfnamefont{J.~D.} \bibnamefont{Bekenstein}}, \bibinfo{journal}{Scholarpedia} \textbf{\bibinfo{volume}{3}}, \bibinfo{pages}{7375} (\bibinfo{year}{2008}), \bibinfo{note}{revision \#182595}.

\bibitem[{\citenamefont{Page}(2005)}]{Page2005}
\bibinfo{author}{\bibfnamefont{D.~N.} \bibnamefont{Page}}, \bibinfo{journal}{New Journal of Physics} \textbf{\bibinfo{volume}{7}}, \bibinfo{pages}{203} (\bibinfo{year}{2005}), \urlprefix\url{https://dx.doi.org/10.1088/1367-2630/7/1/203}.

\bibitem[{\citenamefont{Hossenfelder and Smolin}(2010)}]{HossenfelderSmolin2010}
\bibinfo{author}{\bibfnamefont{S.}~\bibnamefont{Hossenfelder}} \bibnamefont{and} \bibinfo{author}{\bibfnamefont{L.}~\bibnamefont{Smolin}}, \bibinfo{journal}{Phys. Rev. D} \textbf{\bibinfo{volume}{81}}, \bibinfo{pages}{064009} (\bibinfo{year}{2010}), \urlprefix\url{https://link.aps.org/doi/10.1103/PhysRevD.81.064009}.

\bibitem[{\citenamefont{Harlow}(2016)}]{Harlow}
\bibinfo{author}{\bibfnamefont{D.}~\bibnamefont{Harlow}}, \bibinfo{journal}{Rev. Mod. Phys.} \textbf{\bibinfo{volume}{88}}, \bibinfo{pages}{15002} (\bibinfo{year}{2016}).

\bibitem[{\citenamefont{Marolf}(2017)}]{Marolf}
\bibinfo{author}{\bibfnamefont{D.}~\bibnamefont{Marolf}}, \bibinfo{journal}{Rep. Progr. Phys.} \textbf{\bibinfo{volume}{80}}, \bibinfo{pages}{092001} (\bibinfo{year}{2017}).

\bibitem[{\citenamefont{Almheiri et~al.}(2019)\citenamefont{Almheiri, Engelhardt, Marolf, and Maxfield}}]{Almheiri2019}
\bibinfo{author}{\bibfnamefont{A.}~\bibnamefont{Almheiri}}, \bibinfo{author}{\bibfnamefont{N.}~\bibnamefont{Engelhardt}}, \bibinfo{author}{\bibfnamefont{D.}~\bibnamefont{Marolf}}, \bibnamefont{and} \bibinfo{author}{\bibfnamefont{H.}~\bibnamefont{Maxfield}}, \bibinfo{journal}{Journal of High Energy Physics} \textbf{\bibinfo{volume}{2019}}, \bibinfo{pages}{63} (\bibinfo{year}{2019}).

\bibitem[{\citenamefont{Almheiri et~al.}(2020)\citenamefont{Almheiri, Hartman, Maldacena, Shaghoulian, and Tajdini}}]{Almheiri2020}
\bibinfo{author}{\bibfnamefont{A.}~\bibnamefont{Almheiri}}, \bibinfo{author}{\bibfnamefont{T.}~\bibnamefont{Hartman}}, \bibinfo{author}{\bibfnamefont{J.}~\bibnamefont{Maldacena}}, \bibinfo{author}{\bibfnamefont{E.}~\bibnamefont{Shaghoulian}}, \bibnamefont{and} \bibinfo{author}{\bibfnamefont{A.}~\bibnamefont{Tajdini}}, \bibinfo{journal}{Journal of High Energy Physics} \textbf{\bibinfo{volume}{2020}}, \bibinfo{pages}{13} (\bibinfo{year}{2020}).

\bibitem[{\citenamefont{Penington}(2020)}]{Penington2020}
\bibinfo{author}{\bibfnamefont{G.}~\bibnamefont{Penington}}, \bibinfo{journal}{Journal of High Energy Physics} \textbf{\bibinfo{volume}{2020}}, \bibinfo{pages}{2} (\bibinfo{year}{2020}).

\bibitem[{\citenamefont{Almheiri et~al.}(2021)\citenamefont{Almheiri, Hartman, Maldacena, Shaghoulian, and Tajdini}}]{Almeiri21}
\bibinfo{author}{\bibfnamefont{A.}~\bibnamefont{Almheiri}}, \bibinfo{author}{\bibfnamefont{T.}~\bibnamefont{Hartman}}, \bibinfo{author}{\bibfnamefont{J.}~\bibnamefont{Maldacena}}, \bibinfo{author}{\bibfnamefont{E.}~\bibnamefont{Shaghoulian}}, \bibnamefont{and} \bibinfo{author}{\bibfnamefont{A.}~\bibnamefont{Tajdini}}, \bibinfo{journal}{Rev. Mod. Phys.} \textbf{\bibinfo{volume}{93}}, \bibinfo{pages}{035002} (\bibinfo{year}{2021}), \urlprefix\url{https://link.aps.org/doi/10.1103/RevModPhys.93.035002}.

\bibitem[{\citenamefont{Penington et~al.}(2022)\citenamefont{Penington, Shenker, Stanford, and Yang}}]{Penington2022}
\bibinfo{author}{\bibfnamefont{G.}~\bibnamefont{Penington}}, \bibinfo{author}{\bibfnamefont{S.~H.} \bibnamefont{Shenker}}, \bibinfo{author}{\bibfnamefont{D.}~\bibnamefont{Stanford}}, \bibnamefont{and} \bibinfo{author}{\bibfnamefont{Z.}~\bibnamefont{Yang}}, \bibinfo{journal}{Journal of High Energy Physics} \textbf{\bibinfo{volume}{2022}}, \bibinfo{pages}{205} (\bibinfo{year}{2022}).

\bibitem[{\citenamefont{Balasubramanian et~al.}(2024{\natexlab{a}})\citenamefont{Balasubramanian, Lawrence, Mag\'an, and Sasieta}}]{Balasubramanian2024b}
\bibinfo{author}{\bibfnamefont{V.}~\bibnamefont{Balasubramanian}}, \bibinfo{author}{\bibfnamefont{A.}~\bibnamefont{Lawrence}}, \bibinfo{author}{\bibfnamefont{J.~M.} \bibnamefont{Mag\'an}}, \bibnamefont{and} \bibinfo{author}{\bibfnamefont{M.}~\bibnamefont{Sasieta}}, \bibinfo{journal}{Phys. Rev. Lett.} \textbf{\bibinfo{volume}{132}}, \bibinfo{pages}{141501} (\bibinfo{year}{2024}{\natexlab{a}}), \urlprefix\url{https://link.aps.org/doi/10.1103/PhysRevLett.132.141501}.

\bibitem[{\citenamefont{Balasubramanian et~al.}(2024{\natexlab{b}})\citenamefont{Balasubramanian, Lawrence, Mag\'an, and Sasieta}}]{Balasubramanian2024a}
\bibinfo{author}{\bibfnamefont{V.}~\bibnamefont{Balasubramanian}}, \bibinfo{author}{\bibfnamefont{A.}~\bibnamefont{Lawrence}}, \bibinfo{author}{\bibfnamefont{J.~M.} \bibnamefont{Mag\'an}}, \bibnamefont{and} \bibinfo{author}{\bibfnamefont{M.}~\bibnamefont{Sasieta}}, \bibinfo{journal}{Phys. Rev. X} \textbf{\bibinfo{volume}{14}}, \bibinfo{pages}{011024} (\bibinfo{year}{2024}{\natexlab{b}}), \urlprefix\url{https://link.aps.org/doi/10.1103/PhysRevX.14.011024}.

\bibitem[{\citenamefont{Mukhanov}(1986)}]{Mukhanov1986}
\bibinfo{author}{\bibfnamefont{V.}~\bibnamefont{Mukhanov}}, \bibinfo{journal}{Pis. Eksp. Teor. Fiz.} \textbf{\bibinfo{volume}{44}}, \bibinfo{pages}{50} (\bibinfo{year}{1986}), \urlprefix\url{http://jetpletters.ru/ps/1373/article_20808.pdf}.

\bibitem[{\citenamefont{Bekenstein and Mukhanov}(1995)}]{Bekenstein1995}
\bibinfo{author}{\bibfnamefont{J.~D.} \bibnamefont{Bekenstein}} \bibnamefont{and} \bibinfo{author}{\bibfnamefont{V.}~\bibnamefont{Mukhanov}}, \bibinfo{journal}{Physics Letters B} \textbf{\bibinfo{volume}{360}}, \bibinfo{pages}{7} (\bibinfo{year}{1995}), ISSN \bibinfo{issn}{0370-2693}, \urlprefix\url{https://www.sciencedirect.com/science/article/pii/037026939501148J}.

\bibitem[{\citenamefont{Mukhanov}(2018)}]{Mukhanov2018}
\bibinfo{author}{\bibfnamefont{V.}~\bibnamefont{Mukhanov}}, \emph{\bibinfo{title}{Quantum Black Holes}} (\bibinfo{publisher}{World Scientific}, \bibinfo{year}{2018}), pp. \bibinfo{pages}{99--108}.

\bibitem[{\citenamefont{Bekenstein and Meisels}(1977)}]{BekensteinMeisels77}
\bibinfo{author}{\bibfnamefont{J.~D.} \bibnamefont{Bekenstein}} \bibnamefont{and} \bibinfo{author}{\bibfnamefont{A.}~\bibnamefont{Meisels}}, \bibinfo{journal}{Phys. Rev. D} \textbf{\bibinfo{volume}{15}}, \bibinfo{pages}{2775} (\bibinfo{year}{1977}), \urlprefix\url{https://link.aps.org/doi/10.1103/PhysRevD.15.2775}.

\bibitem[{\citenamefont{Hardy and Ramanujan}(1918)}]{HR1918}
\bibinfo{author}{\bibfnamefont{G.~H.} \bibnamefont{Hardy}} \bibnamefont{and} \bibinfo{author}{\bibfnamefont{S.}~\bibnamefont{Ramanujan}}, \bibinfo{journal}{Proceedings of the London Mathematical Society} \textbf{\bibinfo{volume}{2}}, \bibinfo{pages}{75} (\bibinfo{year}{1918}).

\bibitem[{\citenamefont{Erd{\"o}s and Lehner}(1941)}]{EL1941}
\bibinfo{author}{\bibfnamefont{P.}~\bibnamefont{Erd{\"o}s}} \bibnamefont{and} \bibinfo{author}{\bibfnamefont{J.}~\bibnamefont{Lehner}} (\bibinfo{year}{1941}).

\bibitem[{\citenamefont{Auluck and Kothari}(1946)}]{AK1946}
\bibinfo{author}{\bibfnamefont{F.}~\bibnamefont{Auluck}} \bibnamefont{and} \bibinfo{author}{\bibfnamefont{D.}~\bibnamefont{Kothari}}, in \emph{\bibinfo{booktitle}{Mathematical Proceedings of the Cambridge Philosophical Society}} (\bibinfo{organization}{Cambridge University Press}, \bibinfo{year}{1946}), vol.~\bibinfo{volume}{42}, pp. \bibinfo{pages}{272--277}.

\bibitem[{\citenamefont{Agarwala and Auluck}(1951)}]{AA1951}
\bibinfo{author}{\bibfnamefont{B.}~\bibnamefont{Agarwala}} \bibnamefont{and} \bibinfo{author}{\bibfnamefont{F.}~\bibnamefont{Auluck}}, in \emph{\bibinfo{booktitle}{Mathematical Proceedings of the Cambridge Philosophical Society}} (\bibinfo{organization}{Cambridge University Press}, \bibinfo{year}{1951}), vol.~\bibinfo{volume}{47}, pp. \bibinfo{pages}{207--216}.

\bibitem[{\citenamefont{Haselgrove and Temperley}(1954)}]{HT1954}
\bibinfo{author}{\bibfnamefont{C.}~\bibnamefont{Haselgrove}} \bibnamefont{and} \bibinfo{author}{\bibfnamefont{H.}~\bibnamefont{Temperley}}, in \emph{\bibinfo{booktitle}{Mathematical proceedings of the cambridge philosophical society}} (\bibinfo{organization}{Cambridge University Press}, \bibinfo{year}{1954}), vol.~\bibinfo{volume}{50}, pp. \bibinfo{pages}{225--241}.

\bibitem[{\citenamefont{Tran et~al.}(2004)\citenamefont{Tran, Murthy, and Bhaduri}}]{TMB2004}
\bibinfo{author}{\bibfnamefont{M.~N.} \bibnamefont{Tran}}, \bibinfo{author}{\bibfnamefont{M.}~\bibnamefont{Murthy}}, \bibnamefont{and} \bibinfo{author}{\bibfnamefont{R.~K.} \bibnamefont{Bhaduri}}, \bibinfo{journal}{Annals of Physics} \textbf{\bibinfo{volume}{311}}, \bibinfo{pages}{204} (\bibinfo{year}{2004}).

\bibitem[{\citenamefont{Comtet et~al.}(2007{\natexlab{a}})\citenamefont{Comtet, Leboeuf, and Majumdar}}]{CLM2007}
\bibinfo{author}{\bibfnamefont{A.}~\bibnamefont{Comtet}}, \bibinfo{author}{\bibfnamefont{P.}~\bibnamefont{Leboeuf}}, \bibnamefont{and} \bibinfo{author}{\bibfnamefont{S.~N.} \bibnamefont{Majumdar}}, \bibinfo{journal}{Physical review letters} \textbf{\bibinfo{volume}{98}}, \bibinfo{pages}{070404} (\bibinfo{year}{2007}{\natexlab{a}}).

\bibitem[{\citenamefont{Comtet et~al.}(2007{\natexlab{b}})\citenamefont{Comtet, Majumdar, and Ouvry}}]{CMO2007}
\bibinfo{author}{\bibfnamefont{A.}~\bibnamefont{Comtet}}, \bibinfo{author}{\bibfnamefont{S.~N.} \bibnamefont{Majumdar}}, \bibnamefont{and} \bibinfo{author}{\bibfnamefont{S.}~\bibnamefont{Ouvry}}, \bibinfo{journal}{Journal of Physics A: Mathematical and Theoretical} \textbf{\bibinfo{volume}{40}}, \bibinfo{pages}{11255} (\bibinfo{year}{2007}{\natexlab{b}}).

\bibitem[{\citenamefont{Comtet et~al.}(2007{\natexlab{c}})\citenamefont{Comtet, Majumdar, Ouvry, and Sabhapandit}}]{CMOS2007}
\bibinfo{author}{\bibfnamefont{A.}~\bibnamefont{Comtet}}, \bibinfo{author}{\bibfnamefont{S.~N.} \bibnamefont{Majumdar}}, \bibinfo{author}{\bibfnamefont{S.}~\bibnamefont{Ouvry}}, \bibnamefont{and} \bibinfo{author}{\bibfnamefont{S.}~\bibnamefont{Sabhapandit}}, \bibinfo{journal}{Journal of Statistical Mechanics: Theory and Experiment} \textbf{\bibinfo{volume}{2007}}, \bibinfo{pages}{P10001} (\bibinfo{year}{2007}{\natexlab{c}}).

\bibitem[{\citenamefont{Echter et~al.}(2024)\citenamefont{Echter, Maier, Urbina, Lewenkopf, and Richter}}]{EMULR2024}
\bibinfo{author}{\bibfnamefont{C.}~\bibnamefont{Echter}}, \bibinfo{author}{\bibfnamefont{G.}~\bibnamefont{Maier}}, \bibinfo{author}{\bibfnamefont{J.-D.} \bibnamefont{Urbina}}, \bibinfo{author}{\bibfnamefont{C.}~\bibnamefont{Lewenkopf}}, \bibnamefont{and} \bibinfo{author}{\bibfnamefont{K.}~\bibnamefont{Richter}}, \bibinfo{journal}{Journal of Physics A: Mathematical and Theoretical}  (\bibinfo{year}{2024}).

\bibitem[{\citenamefont{Peyen et~al.}(2024)\citenamefont{Peyen, Bogachev, and Martin}}]{PBM2024}
\bibinfo{author}{\bibfnamefont{J.~C.} \bibnamefont{Peyen}}, \bibinfo{author}{\bibfnamefont{L.~V.} \bibnamefont{Bogachev}}, \bibnamefont{and} \bibinfo{author}{\bibfnamefont{P.~P.} \bibnamefont{Martin}}, \bibinfo{journal}{Advances in Applied Mathematics} \textbf{\bibinfo{volume}{159}}, \bibinfo{pages}{102739} (\bibinfo{year}{2024}).

\bibitem[{\citenamefont{Dvali and Gomez}(2011)}]{dvali2011blackholesquantumnportrait}
\bibinfo{author}{\bibfnamefont{G.}~\bibnamefont{Dvali}} \bibnamefont{and} \bibinfo{author}{\bibfnamefont{C.}~\bibnamefont{Gomez}}, \emph{\bibinfo{title}{Black hole's quantum n-portrait}} (\bibinfo{year}{2011}), \eprint{1112.3359}, \urlprefix\url{https://arxiv.org/abs/1112.3359}.

\bibitem[{\citenamefont{Dvali}(2016)}]{Dvali}
\bibinfo{author}{\bibfnamefont{G.}~\bibnamefont{Dvali}}, \bibinfo{journal}{Fortschritte der Physik} \textbf{\bibinfo{volume}{64}}, \bibinfo{pages}{106} (\bibinfo{year}{2016}), \eprint{https://onlinelibrary.wiley.com/doi/pdf/10.1002/prop.201500096}, \urlprefix\url{https://onlinelibrary.wiley.com/doi/abs/10.1002/prop.201500096}.

\bibitem[{\citenamefont{Poisson and Israel}(1989)}]{Poisson1989}
\bibinfo{author}{\bibfnamefont{E.}~\bibnamefont{Poisson}} \bibnamefont{and} \bibinfo{author}{\bibfnamefont{W.}~\bibnamefont{Israel}}, \bibinfo{journal}{Phys. Rev. Lett.} \textbf{\bibinfo{volume}{63}}, \bibinfo{pages}{1663} (\bibinfo{year}{1989}), \urlprefix\url{https://link.aps.org/doi/10.1103/PhysRevLett.63.1663}.

\bibitem[{\citenamefont{Bonanno et~al.}(1995)\citenamefont{Bonanno, Droz, Israel, and Morsink}}]{Bonnanno1995}
\bibinfo{author}{\bibfnamefont{A.}~\bibnamefont{Bonanno}}, \bibinfo{author}{\bibfnamefont{S.}~\bibnamefont{Droz}}, \bibinfo{author}{\bibfnamefont{W.}~\bibnamefont{Israel}}, \bibnamefont{and} \bibinfo{author}{\bibfnamefont{S.~M.} \bibnamefont{Morsink}}, \bibinfo{journal}{Proceedings of the Royal Society of London. Series A: Mathematical and Physical Sciences} \textbf{\bibinfo{volume}{450}}, \bibinfo{pages}{553} (\bibinfo{year}{1995}), \eprint{https://royalsocietypublishing.org/doi/pdf/10.1098/rspa.1995.0100}, \urlprefix\url{https://royalsocietypublishing.org/doi/abs/10.1098/rspa.1995.0100}.

\bibitem[{\citenamefont{Christodoulou}(1999)}]{Christodoulou1999}
\bibinfo{author}{\bibfnamefont{D.}~\bibnamefont{Christodoulou}}, \bibinfo{journal}{Classical and Quantum Gravity} \textbf{\bibinfo{volume}{16}}, \bibinfo{pages}{A23} (\bibinfo{year}{1999}), \urlprefix\url{https://dx.doi.org/10.1088/0264-9381/16/12A/302}.

\bibitem[{\citenamefont{Dafermos}(2005)}]{Dafermos2005}
\bibinfo{author}{\bibfnamefont{M.}~\bibnamefont{Dafermos}}, \bibinfo{journal}{Communications on Pure and Applied Mathematics} \textbf{\bibinfo{volume}{58}}, \bibinfo{pages}{445} (\bibinfo{year}{2005}), \eprint{https://onlinelibrary.wiley.com/doi/pdf/10.1002/cpa.20071}, \urlprefix\url{https://onlinelibrary.wiley.com/doi/abs/10.1002/cpa.20071}.

\bibitem[{\citenamefont{Dafermos and Luk}(2017)}]{dafermos2017interiordynamicalvacuumblack}
\bibinfo{author}{\bibfnamefont{M.}~\bibnamefont{Dafermos}} \bibnamefont{and} \bibinfo{author}{\bibfnamefont{J.}~\bibnamefont{Luk}}, \emph{\bibinfo{title}{The interior of dynamical vacuum black holes i: The $c^0$-stability of the kerr cauchy horizon}} (\bibinfo{year}{2017}), \eprint{1710.01722}, \urlprefix\url{https://arxiv.org/abs/1710.01722}.

\bibitem[{\citenamefont{De~Clerck et~al.}(2024)\citenamefont{De~Clerck, Hartnoll, and Santos}}]{DeClerck2024}
\bibinfo{author}{\bibfnamefont{M.}~\bibnamefont{De~Clerck}}, \bibinfo{author}{\bibfnamefont{S.~A.} \bibnamefont{Hartnoll}}, \bibnamefont{and} \bibinfo{author}{\bibfnamefont{J.~E.} \bibnamefont{Santos}}, \bibinfo{journal}{Journal of High Energy Physics} \textbf{\bibinfo{volume}{2024}}, \bibinfo{pages}{202} (\bibinfo{year}{2024}).

\bibitem[{\citenamefont{Bueno et~al.}(2024)\citenamefont{Bueno, Cano, and Hennigar}}]{Bueno2024}
\bibinfo{author}{\bibfnamefont{P.}~\bibnamefont{Bueno}}, \bibinfo{author}{\bibfnamefont{P.~A.} \bibnamefont{Cano}}, \bibnamefont{and} \bibinfo{author}{\bibfnamefont{R.~A.} \bibnamefont{Hennigar}}, \bibinfo{journal}{Phys. Rev. D} \textbf{\bibinfo{volume}{110}}, \bibinfo{pages}{L041503} (\bibinfo{year}{2024}), \urlprefix\url{https://link.aps.org/doi/10.1103/PhysRevD.110.L041503}.

\bibitem[{\citenamefont{C{\'a}ceres et~al.}(2024)\citenamefont{C{\'a}ceres, Murcia, Patra, and Pedraza}}]{Caceres2024}
\bibinfo{author}{\bibfnamefont{E.}~\bibnamefont{C{\'a}ceres}}, \bibinfo{author}{\bibfnamefont{{\'A}.~J.} \bibnamefont{Murcia}}, \bibinfo{author}{\bibfnamefont{A.~K.} \bibnamefont{Patra}}, \bibnamefont{and} \bibinfo{author}{\bibfnamefont{J.~F.} \bibnamefont{Pedraza}}, \bibinfo{journal}{Journal of High Energy Physics} \textbf{\bibinfo{volume}{2024}}, \bibinfo{pages}{77} (\bibinfo{year}{2024}).

\bibitem[{\citenamefont{Zilberman et~al.}(2022)\citenamefont{Zilberman, Casals, Ori, and Ottewill}}]{Zilberman2022}
\bibinfo{author}{\bibfnamefont{N.}~\bibnamefont{Zilberman}}, \bibinfo{author}{\bibfnamefont{M.}~\bibnamefont{Casals}}, \bibinfo{author}{\bibfnamefont{A.}~\bibnamefont{Ori}}, \bibnamefont{and} \bibinfo{author}{\bibfnamefont{A.~C.} \bibnamefont{Ottewill}}, \bibinfo{journal}{Phys. Rev. Lett.} \textbf{\bibinfo{volume}{129}}, \bibinfo{pages}{261102} (\bibinfo{year}{2022}), \urlprefix\url{https://link.aps.org/doi/10.1103/PhysRevLett.129.261102}.

\bibitem[{\citenamefont{Carballo-Rubio et~al.}(2024)\citenamefont{Carballo-Rubio, Di~Filippo, Liberati, and Visser}}]{Carballo-Rubio2024}
\bibinfo{author}{\bibfnamefont{R.}~\bibnamefont{Carballo-Rubio}}, \bibinfo{author}{\bibfnamefont{F.}~\bibnamefont{Di~Filippo}}, \bibinfo{author}{\bibfnamefont{S.}~\bibnamefont{Liberati}}, \bibnamefont{and} \bibinfo{author}{\bibfnamefont{M.}~\bibnamefont{Visser}}, \bibinfo{journal}{Phys. Rev. Lett.} \textbf{\bibinfo{volume}{133}}, \bibinfo{pages}{181402} (\bibinfo{year}{2024}), \urlprefix\url{https://link.aps.org/doi/10.1103/PhysRevLett.133.181402}.

\bibitem[{\citenamefont{Wheeler}(1963)}]{Wheeler1963}
\bibinfo{author}{\bibfnamefont{J.}~\bibnamefont{Wheeler}}, \emph{\bibinfo{title}{Geometrodynamics and the Issue of the Final State}} (\bibinfo{publisher}{{Gordon and Breach}}, \bibinfo{year}{1963}).

\bibitem[{\citenamefont{Chakravarty}(2021)}]{Chakravarty2021}
\bibinfo{author}{\bibfnamefont{J.}~\bibnamefont{Chakravarty}}, \bibinfo{journal}{Journal of High Energy Physics} \textbf{\bibinfo{volume}{2021}}, \bibinfo{pages}{27} (\bibinfo{year}{2021}).

\bibitem[{\citenamefont{M\'ezard and Montanari}(2009)}]{MezardMontanari2009}
\bibinfo{author}{\bibfnamefont{M.}~\bibnamefont{M\'ezard}} \bibnamefont{and} \bibinfo{author}{\bibfnamefont{A.}~\bibnamefont{Montanari}}, \emph{\bibinfo{title}{Information, Physics, and Computation}} (\bibinfo{publisher}{Oxford University Press}, \bibinfo{year}{2009}), ISBN \bibinfo{isbn}{9780198570837}, \urlprefix\url{https://doi.org/10.1093/acprof:oso/9780198570837.001.0001}.

\bibitem[{\citenamefont{Charbonneau et~al.}(2023)\citenamefont{Charbonneau, Marinari, M\'ezard, Parisi, Ricci-Tersenghi, Sicuro, and Zamponi}}]{FarBeyond}
\bibinfo{author}{\bibfnamefont{P.}~\bibnamefont{Charbonneau}}, \bibinfo{author}{\bibfnamefont{E.}~\bibnamefont{Marinari}}, \bibinfo{author}{\bibfnamefont{M.}~\bibnamefont{M\'ezard}}, \bibinfo{author}{\bibfnamefont{G.}~\bibnamefont{Parisi}}, \bibinfo{author}{\bibfnamefont{F.}~\bibnamefont{Ricci-Tersenghi}}, \bibinfo{author}{\bibfnamefont{G.}~\bibnamefont{Sicuro}}, \bibnamefont{and} \bibinfo{author}{\bibfnamefont{F.}~\bibnamefont{Zamponi}}, \emph{\bibinfo{title}{Spin Glass Theory and Far Beyond}} (\bibinfo{publisher}{WORLD SCIENTIFIC}, \bibinfo{year}{2023}), \eprint{https://www.worldscientific.com/doi/pdf/10.1142/13341}, \urlprefix\url{https://www.worldscientific.com/doi/abs/10.1142/13341}.

\bibitem[{\citenamefont{M\'ezard and Parisi}(2003)}]{MezardParisi2003}
\bibinfo{author}{\bibfnamefont{M.}~\bibnamefont{M\'ezard}} \bibnamefont{and} \bibinfo{author}{\bibfnamefont{G.}~\bibnamefont{Parisi}}, \bibinfo{journal}{Journal of Statistical Physics} \textbf{\bibinfo{volume}{111}}, \bibinfo{pages}{1} (\bibinfo{year}{2003}).

\end{thebibliography}

\end{document}